\documentclass[9pt,twoside]{article}
\usepackage{graphicx}
\usepackage{amsmath}

\begin{document}

{\noindent\bf \Large Deterministic secure direct communication using GHZ states and swapping quantum entanglement}\\[0.2cm]

\leftskip 2cm

{\noindent\bf T Gao$^{1,2,3}$,  F L Yan$^{1,4}$ and Z X Wang$^3$}\\[0.2cm]

 {\noindent\footnotesize $^1$ CCAST (World Laboratory), P.O. Box 8730, Beijing 100080, People's Republic of China\\
$^2$ College of Mathematics and Information Science, Hebei Normal University, Shijiazhuang 050016,
People's Republic of China\\
 $^3$ Department of Mathematics, Capital Normal University, Beijing 100037, People's Republic of China\\
$^4$ College of Physics, Hebei Normal University, Shijiazhuang 050016, People's Republic of China\\
[0.2cm]
E-mail: gaoting@heinfo.net\\[0.2cm]}
{{\bf Abstract}\\
\noindent We present a deterministic secure direct communication  scheme via entanglement swapping, where a set
of ordered maximally entangled three-particle states (GHZ states), initially shared by three spatially separated
parties, Alice,  Bob and Charlie,  functions as a quantum information channel. After ensuring the safety of the
quantum channel, Alice and Bob apply a series local operations on their  respective particles according to the
tripartite stipulation  and the secret message they both want to send to Charlie. By three Alice, Bob and
Charlie's Bell measurement results, Charlie is able to infer the secret messages directly.  The secret messages
are faithfully transmitted from Alice and Bob to Charlie via initially shared pairs of GHZ states without
revealing any information to a potential eavesdropper. Since there is not a transmission of the qubits
 carrying the secret message between any two of them in the public channel, it is
completely secure for direct secret communication if perfect quantum channel is used.

\noindent PACS numbers: 03.67.Dd, 03.67.Hk\\[0.5cm]}

\leftskip 0cm

{\noindent{\bf 1. Introduction  }}\\[0.2cm]
Cryptography is the art of enabling two parties to communicate in private. Effective cryptosystems make it easy
for parties who wish to communicate to achieve privacy, but make it very difficult for third parties to
`eavesdrop' on the content of the conversation. A simple, yet highly effective private key cryptosystem is the
{\it Vernam cipher}, sometimes called a one time pad. The great feature of this system is that as long as the
key strings are truly secret, it is provably secure. The major difficulty of private key cryptosystems is secure
distribution of key bits, since a malevolent third party may be eavesdropping on the key distribution, and then
use the intercepted key to decrypt some of the message in transmission.

One of the earliest discoveries in quantum computation and quantum information was that quantum mechanics can be
used to do key distribution in such a way that Alice and Bob's security can not be compromised. This procedure
is known as {\it quantum cryptography or quantum key distribution} (QKD). The basic idea is to exploit the
quantum mechanical principle that observation in general disturbs the system being observed. In 1984, Bennett
and Brassard proposed the first quantum cryptography  protocol \cite {BB84} using quantum mechanics to
distribute keys between Alice and Bob, without any possibility of a compromise. Since then numerous QKD
protocols have been proposed, such as Ekert 1991 protocol (Ekert91) \cite {Ekert91}, Bennett-Brassard-Mermin
1992 protocol (BBM92) \cite {BBM92}, B92 protocol \cite {B92} and other protocols \cite {BW, GV, HIGM, KI, BBBW,
Bru, HKH, Cabelloprl, Cabellopra, LL, XLG, PBTB, LCA, Song, Wang, HBB, BTL, IWY}.

Recently, Shimizu and Imoto \cite {SIpra60, SIpra62} and Beige et al. \cite {Beige} presented  novel quantum
secure direct communication (QSDC) schemes, in which the two parties communicate important messages directly
without first establishing a shared secret key to encrypt them and the message is deterministically sent through
the quantum channel, but can be read only after a transmission of an additional  classical information for each
qubit. Bostr\"{o}m and Felbinger \cite {BF} put forward a communication scheme, the "ping-pong protocol", which
also allows for deterministic communication. This protocol can be used for the transmission of either a secret
key or a plaintext message. W\'{o}jcik  discussed the security of
   the "ping-pong protocol" in a noisy quantum channel  \cite {Wojcik}. Deng et al. \cite {DLL} suggested a two-step
quantum direct communication protocol using Einstein-Podolsky-Rosen pair block. However in all these QSDC
schemes it is necessary to send the qubits with  secret messages (message-coding sequence) in the public
channel. Therefore, Eve can attack the qubits  in transmission and  make the communication interrupt.

 More recently, Yan and Zhang \cite {YZ} presented  a QSDC scheme
using  Einstein-Podolsky-Rosen pairs and teleportation  \cite
{BBCJPW93}. By means of controlled quantum teleportation \cite
{KB} we proposed two controlled QSDC protocols \cite {Gaozna,
GYWcp}. Since in these protocols there are no particles carrying
secret messages to be transmitted in public channel,  so the
communication can not be interrupted by any eavesdropper.
Therefore, they are completely secure for direct secret
communication as long as perfect quantum channel is used.

Entanglement swapping \cite {ZZHE} is a method that enables one to entangle two quantum systems that do not have
direct interaction with one another. Based on entanglement swapping, we presented a QSDC scheme \cite {GYWnc}.
In this paper, we introduce another QSDC scheme achieved by swapping quantum entanglement, in which a set of
ordered three-particle Greenberger-Horne-Zeilinger (GHZ) states initially shared by three spatially separated
parties, Alice,  Bob and Charlie,  functions as a quantum information channel.  The proposed
 QSDC scheme   is simultaneous mutual  communications among different pairs of parties, one for Alice and Charlie
 and another for Bob and  Charlie.  After ensuring the safety of the quantum channel, Alice and Bob encode
 secret classical bits by applying predetermined unitary operations on GHZ triplets. The
   secret messages encoded by local operations are faithfully transmitted
from two distant senders (Alice and Bob) to a remote receiver (Charlie)  without revealing any information to a
potential eavesdropper. \\[0.2cm]
{\noindent\bf 2. A simultaneous mutual  quantum secure direct communication protocol between the central party
and other two parties}\\[0.2cm]
In this section we propose a simultaneous mutual  quantum secure direct communication scheme which utilizes
shared GHZ states and entanglement swapping between communicating parties, in the form three people (Alice, Bob
and Charlie).\\[0.2cm]
{\it 2.1 Notation}\\[0.2cm]
  Let us start by illustrating entanglement swapping. We first define four Bell states (EPR pairs) as
\begin{equation}\label{2EPR}
    \Phi^{\pm}\equiv\frac{1}{\sqrt{2}}(|00\rangle\pm |11\rangle),
    \Psi^{\pm}\equiv\frac{1}{\sqrt{2}}(|01\rangle\pm|10\rangle),
\end{equation}
and eight GHZ states as
\begin{equation}\label{2GHZ}
  \begin{array}{cc}
  |P^{\pm}\rangle\equiv\frac{1}{\sqrt{2}}(|000\rangle\pm|111\rangle),  &
    |Q^{\pm}\rangle\equiv\frac{1}{\sqrt{2}}(|001\rangle\pm|110\rangle),\\
  |R^{\pm}\rangle\equiv\frac{1}{\sqrt{2}}(|010\rangle\pm|101\rangle),  &
    |S^{\pm}\rangle\equiv\frac{1}{\sqrt{2}}(|011\rangle\pm|100\rangle).
  \end{array}
\end{equation}
Suppose three  parties, Alice, Bob and Charlie, share two GHZ triplets $|P^+\rangle_{123}$ and
$|P^+\rangle_{456}$ where Alice has qubits 1 and 4,  Bob possesses 2 and 5, and particles 3 and 6 are in
Charlie's side. Two operations are performed on qubits 1 and 4, and 2 and 5  with the Bell basis, $\Phi^{\pm}$
and $\Psi^{\pm}$, by Alice and Bob, respectively,  then the total state
$|P^+\rangle_{123}\otimes|P^+\rangle_{456}$ is projected onto $\Phi^+_{14}\otimes\Phi^+_{25}\otimes\Phi^+_{36}$,
$\Phi^+_{14}\otimes\Phi^-_{25}\otimes\Phi^-_{36}$, $\Phi^-_{14}\otimes\Phi^+_{25}\otimes\Phi^-_{36}$,
$\Phi^-_{14}\otimes\Phi^-_{25}\otimes\Phi^+_{36}$, $\Psi^+_{14}\otimes\Psi^+_{25}\otimes\Psi^+_{36}$,
$\Psi^+_{14}\otimes\Psi^-_{25}\otimes\Psi^-_{36}$, $\Psi^-_{14}\otimes\Psi^+_{25}\otimes\Psi^-_{36}$ and
$\Psi^-_{14}\otimes\Psi^-_{25}\otimes\Psi^+_{36}$ with equal probability of 1/8 for each. Previous entanglement
of qubits 1, 2 and 3, and 4, 5 and 6 are now swapped into entanglement between 1 and 4, 2 and 5, and 3 and 6.
Although we considered entanglement swapping with the initial state $|P^+\rangle_{123}\otimes|P^+\rangle_{456}$,
similar results can be achieved with other GHZ states. For example, when Alice, Bob and Charlie originally share
$|S^+\rangle_{123}$ and $|P^+\rangle_{456}$, there are eight possible measurement outcomes,
$\Psi^+_{14}\otimes\Phi^+_{25}\otimes\Phi^+_{36}$, $\Psi^+_{14}\otimes\Phi^-_{25}\otimes\Phi^-_{36}$,
$\Psi^-_{14}\otimes\Phi^+_{25}\otimes\Phi^-_{36}$, $\Psi^-_{14}\otimes\Phi^-_{25}\otimes\Phi^+_{36}$,
$\Phi^+_{14}\otimes\Psi^+_{25}\otimes\Psi^+_{36}$, $\Phi^+_{14}\otimes\Psi^-_{25}\otimes\Psi^-_{36}$,
$\Phi^-_{14}\otimes\Psi^+_{25}\otimes\Psi^-_{36}$ and $\Phi^-_{14}\otimes\Psi^-_{25}\otimes\Psi^+_{36}$ with
equal probability  1/8.\\[0.2cm]
{\it 2.2 Preparing quantum channel}\\[0.2cm]
Suppose that three spatially separated  parties wish to realize  simultaneous mutual communications in secret
among different pairs of parties, one for Alice and Charlie and another for Bob and  Charlie.  In order to
achieve tripartite communications between one party and the other two parties in private at the same time, the
first step  is to establish quantum channel (GHZ triplets). Obtaining these GHZ triplets could have come about
in many different ways, such as  Charlie prepares a sequence of GHZ triplets and then share each  triplet with
Alice and Bob;  or they could have met
 a long time ago and shared them,
 storing them until the present.  Alice,  Bob and Charlie then choose  randomly a subset of GHZ triplets,
  do   some  appropriate tests of fidelity.  Passing the test certifies that
  they continue to hold sufficiently pure, entangled quantum states.  However,
 if tampering has occurred,
 they throw out the GHZ triplets and reconstruct them. We will discuss the details  in  Section
 3.\\[0.2cm]
{\it 2.3 The direct communication scheme by shared GHZ states and entanglement swapping }\\[0.2cm]
After ensuring the security of the quantum channel (GHZ states), Alice, Bob and Charlie begin
  secure direct communication. The  QSDC scheme works as follows:

 (1)  Alice, Bob and Charlie  randomly divide all pure GHZ triplets into $N$ ordered groups
$\{\xi(1)_{123},$ $ \eta(1)_{456}\}$, $\{\xi(2)_{123},$ $\eta(2)_{456}\}$, $\cdots$, $\{\xi(N)_{123},$
$\eta(N)_{456}\}$,
 where    $\xi(i)_{123}$ and $\eta(i)_{456}$ denote two GHZ states of Alice's particles 1 and 4,  Bob's particles
  2 and 5, and Charlie's 3 and 6 in the  $i$-th group.  For simplicity, let us suppose that these GHZ triplets are in the state
$|P^+\rangle$.

(2) Alice, Bob and Charlie agree on  that Alice encodes information by  local operations
\begin{equation}\label{2operation1}
\begin{array}{cc}
 \sigma_{00}=I=|0\rangle\langle 0|+|1\rangle\langle1|, & \sigma_{01}=\sigma_x=|0\rangle\langle1|+|1\rangle\langle0|, \\
 \sigma_{10}={\rm i}\sigma_y=|0\rangle\langle1|-|1\rangle\langle0|, &
\sigma_{11}=\sigma_z=|0\rangle\langle0|-|1\rangle\langle1|
\end{array}
\end{equation}
on GHZ triplets $\xi(i)_{123}$, and  Bob by local operations
\begin{equation}\label{2operation2}
 \sigma_0=I=|0\rangle\langle 0|+|1\rangle\langle1|, ~~ \sigma_1=\sigma_x=|0\rangle\langle1|+|1\rangle\langle0|.
\end{equation}
 Alice and Charlie, and Bob and Charlie  assign secretely two bits and one bit to Alice and Bob's respective operations
 as following encoding
\begin{equation}\label{2m12}
\sigma_{00}\rightarrow 00,~~ \sigma_{01}\rightarrow 01, ~~\sigma_{10}\rightarrow 10, ~~\sigma_{11}\rightarrow
11,
\end{equation}
and
\begin{equation}\label{2m11}
 \sigma_0\rightarrow 0,~~ \sigma_1\rightarrow 1.
\end{equation}

(3) Alice and Bob  encode their respective messages (secret classical bits) on  GHZ groups. Explicitly, both
Alice and Bob apply a predetermined unitary operation on each of their particles 1 and 2 according to their
respective secret message sequence, respectively.

 Suppose Alice, Bob, and Charlie initially share GHZ state $|P^+\rangle_{123}, |P^+\rangle_{456}$,
  then the originally total state of them is
\begin{eqnarray}\label{2original}
 &&|P^+\rangle_{123}\otimes|P^+\rangle_{456}\nonumber\\
&=&\frac{1}{2\sqrt{2}}[\Phi^+_{14}\otimes\Phi^+_{25}\otimes\Phi^+_{36}+\Phi^+_{14}\otimes\Phi^-_{25}\otimes\Phi^-_{36}
 +\Phi^-_{14}\otimes\Phi^+_{25}\otimes\Phi^-_{36}+\Phi^-_{14}\otimes\Phi^-_{25}\otimes\Phi^+_{36}\nonumber\\
 &&
 +\Psi^+_{14}\otimes\Psi^+_{25}\otimes\Psi^+_{36}+\Psi^+_{14}\otimes\Psi^-_{25}\otimes\Psi^-_{36}
 +\Psi^-_{14}\otimes\Psi^+_{25}\otimes\Psi^-_{36}+\Psi^-_{14}\otimes\Psi^-_{25}\otimes\Psi^+_{36}].
\end{eqnarray}
If Alice wishes to transmit 11 to Charlie and Bob wants to send 1 to Charlie,  then Alice  performs a local
operation $\sigma_{11}$ on particle 1 and Bob applies $\sigma_1$ on his particle 2, thus  the
   state $|P^+\rangle_{123}$ is turned into $|R^-\rangle_{123}$.

(4) Alice and Bob make a Bell measurement on  particles 1 and 4, and 2 and 5, respectively. We can see the
effects of measurements by Alice and Bob on Charlie's particles 3 and 6 if we express the product of GHZ states
$|R^-\rangle_{123}$ and $|P^+\rangle_{456}$ in the following equation:
\begin{eqnarray}\label{2factual}
&&|R^-\rangle_{123}\otimes|P^+\rangle_{456}\nonumber\\
&=&\frac{1}{2\sqrt{2}}[\Phi^-_{14}\otimes\Psi^+_{25}\otimes\Phi^+_{36}-\Phi^-_{14}\otimes\Psi^-_{25}\otimes\Phi^-_{36}
 +\Phi^+_{14}\otimes\Psi^+_{25}\otimes\Phi^-_{36}-\Phi^+_{14}\otimes\Psi^-_{25}\otimes\Phi^+_{36}\nonumber\\
 &&+\Psi^-_{14}\otimes\Phi^+_{25}\otimes\Psi^+_{36}
 -\Psi^-_{14}\otimes\Phi^-_{25}\otimes\Psi^-_{36}
 +\Psi^+_{14}\otimes\Phi^+_{25}\otimes\Psi^-_{36}-\Psi^+_{14}\otimes\Phi^-_{25}\otimes\Psi^+_{36}].
\end{eqnarray}
If  Alice and Bob get  measurement outcomes  $\Phi^+_{14}$ and $\Psi^-_{25}$, respectively, then Charlie's two
particles 3 and 6 will have the state $\Phi^+_{36}$.

(5) Alice and Bob inform Charlie that they have made a Bell measurement on  particles 1 and 4, and 2 and 5 over
a classical channel, respectively, but do not tell the results of their measurements.

(6) Charlie performs a Bell measurement on his particles 3 and 6 and  deduces the outcomes of both Alice and
Bob's measurements.

 From the calculation of entanglement swapping (Eq.(\ref{2original}))  and his measurement outcome $\Phi^+_{36}$,
 Charlie could calculate  that the initially whole state $|P^+\rangle_{123}\otimes|P^+\rangle_{456}$ should collapse to
$\Phi^+_{14}\otimes\Phi^+_{25}\otimes\Phi^+_{36}$ or
 $\Phi^-_{14}\otimes\Phi^-_{25}\otimes\Phi^+_{36}$   without Alice and Bob's local operations.

(7) Charlie asks and gets Alice and Bob's measurement results publicly.

(8) Charlie can read out Alice and Bob's secret message by comparing his calculation result  with Alice and
Bob's practical measurement outcomes.

From the  measurement results announced by Alice and Bob, and his calculation result, Charlie can infer that
Alice and Bob  have applied  local operations $\sigma_{11}$ and $\sigma_1$ on particles 1 and 2, respectively,
such that $\Phi^-_{14}\otimes\Phi^-_{25}\otimes\Phi^+_{36}$ turns into
$\Phi^+_{14}\otimes\Psi^-_{25}\otimes\Phi^+_{36}$, since it is impossible for Alice and Bob to change
$\Phi^+_{14}\otimes\Phi^+_{25}\otimes\Phi^+_{36}$ into $\Phi^+_{14}\otimes\Psi^-_{25}\otimes\Phi^+_{36}$ by
applying unitary operation $\sigma_{k_1k_{1'}}\otimes\sigma_{k_2}$ ($k_1, k_{1'}, k_2 \in \{0, 1\}$) on
particles 1 and 2, thus he obtains Alice's message 11 and Bob's 1. Finally,  three spatially separated parties
have realized deterministic secure direct communication between one party and  the other two parties.

Remark 1:   We should point out that the  encoding schemes of Eq.(\ref{2m12}) and Eq.(\ref{2m11}) are secrete,
i.e. only Alice and Charlie  know the encoding scheme Eq.(\ref{2m12}),   and only Bob and Charlie know
Eq.(\ref{2m11}). The reason is as follows. After Charlie performs a Bell measurement on his qubits 3 and 6, he
asks Alice and Bob to declare their Bell measurement results on the qubits 1 and 4, and 2 and 5. This public
declaration step is crucial. However, an eavesdropper who knows that the original initial state is
$|P^+\rangle_{123}\otimes|P^+\rangle_{456}$ will do her calculation the same as Charlie. When she hears that,
Alice and Bob, respectively, obtained measurement results $\Phi^+_{14}$ and $\Psi^-_{25}$, the eavesdropper
looks at Eq.(\ref{2original}) and can easily deduce that such a measurement result can be obtained by applying
one-qubit unitary operators on the following four cases: $\Phi^+_{14}\otimes\Phi^-_{25}\otimes\Phi^-_{36}$,
$\Phi^-_{14}\otimes\Phi^-_{25}\otimes\Phi^+_{36}$, $\Psi^+_{14}\otimes\Psi^-_{25}\otimes\Psi^-_{36}$, and
$\Psi^-_{14}\otimes\Psi^-_{25}\otimes\Psi^+_{36}$. Since Charlie's measurement result is secret (not publicly
declared), the eavesdropper may pick the correct state only with a probability of 1/4. If the information on the
encoding scheme is not available to the eavesdropper, there is no way for the eavesdropper to find the correct
classical bits. So it is necessary for the two pairs, Alice and Charlie, and Bob and Charlie,  to keep their
respective encoding schemes Eq.(\ref{2m12}) and  Eq.(\ref{2m11})  in private. In order to achieve privacy
safely, Alice and Charlie, and Bob and Charlie may use secret keys generated by shared GHZ states  to
communicate the encoding method with each other. Since Alice, Bob and Charlie want to achieve simultaneous
mutual communications in secret among different pairs of parties, one for Alice and Charlie and another for Bob
and  Charlie, they three must be trustworthy and cooperative. Two communication parties Alice and Charlie (Bob
and Charlie) can generate secret key used to transmit their encoding scheme via initially shared pairs of GHZ
states with the help of the third party Bob (Alice). The details of generating secret keys are as follows.
Suppose Alice wants to send  Charlie her encoding scheme. Each of Alice, Bob and Charlie performs a Bell
measurement on their respective particles 1 and 4,  2 and 5, and 3 and 6,  obtaining one of four possible
results, $\Phi^+$, $\Phi^-$, $\Psi^+$ and $\Psi^-$. Bob tells Alice and Charlie of his measurement outcome.
Depending on Bob's information, Alice and Charlie can infer the measurement result of each other. Alice and
Charlie agree on that each of the four Bell states carry two bits classical message (there are 4!=24 kinds of
encoding methods, they can choose one kind at random) and regard the information carrying by either Alice's
measurement results or Bob's measurement results as secret key bits used to transmit their encoding scheme. For
instance, if the original state is $|R^-\rangle_{123}|P^+\rangle_{456}$, and the outcome of Bob's measurement is
$\Psi^+_{25}$, then according to her measurement result $\Phi^-_{14}$, Alice can infer that the outcome of
Charlie's measurement must be $\Phi^+_{36}$. Similarly, Charlie can deduce Alice's measurement result
$\Phi^-_{14}$ from his measurement outcome $\Phi^+_{36}$. If Alice and Charlie encode $\Phi^+_{14}$,
$\Phi^-_{14}$, $\Psi^+_{14}$, and $\Psi^-_{14}$ as 00, 01, 10, and 11, then they share two classical bits 01.
Alice and Charlie  sacrifice some randomly selected bits to test the ``error rate". If the error rate is too
high, they abort this QKD protocol. Otherwise, they perform information reconciliation and privacy amplification
\cite{MW, BBCM, Hamming, BS, BLTNDP, Pearson, ACC} on the remaining bits to obtain secure final key bits for
Alice informing Charlie of the encoding scheme Eq.(\ref{2m12}).  Using the same method, Bob and Charlie get
secret key for Bob sending the encoding scheme Eq.(\ref{2m11}) to Charlie.
 Thus,
 in our QSDC scheme, the eavesdropper can not get the encoding scheme of the
classical bits on the unitary operators.  That is, if the eavesdropper understands that the operator Alice has
applied is $\sigma_z$, she  does not know that this corresponds to the classical bits 11. Therefore,  our scheme
is a deterministic QSDC scheme.

Remark 2: The crucial point in the proposed scheme is that the qubits carrying the encoded message are not
transmitted in the public channel. Therefore, a potential eavesdropper cannot obtain any information.

Remark 3: In order to protect the transmitting information from the eavesdropper, Alice, Bob,  and Charlie can
make use of classical error correction protocol \cite{Hamming}. That is, Alice and Bob encode their secret
message and Charlie decode these message according to a pre-determined  classical error correction protocol.

Note: (A) The  above protocol is also a quantum key distribution (QKD) scheme based on GHZ states and
entanglement swapping. If the communication parties want to distribute  keys, Alice and Bob randomly generate
their respective classical bit strings $a$ and $b$, and then Alice divides her string $a$ in length of two bits
and encode by applying the unitary operators on her qubits and in the same way Bob applies his operator for each
classical bit of $b$. Alice, Bob, and Charlie agree upon in advance that each of the four Bell states can carry
two bits classical information and encode $\Phi^+$, $\Phi^-$, $\Psi^+$, and $\Psi^-$  as 00, 01, 10, and 11,
respectively. Protocol then follows before. By Alice's measurement result $\Phi^+_{14}$, only both Alice and
Charlie derive $\Phi^-_{14} \stackrel{\sigma_{11}}{\longrightarrow} \Phi^+_{14}$, i.e. Alice and Charlie obtain
 $\sigma_{11}$ and $\Phi^-_{14}$ secretly. Since Alice's operator $\sigma_{11}$ is certain and her measurement
result $\Phi^+_{14}$ is random,  Alice and Charlie share two certain bits 11 and two random bits 01 in private.
Similarly, from Bob's measurement outcome $\Psi^-_{25}$, Bob and Charlie obtain $\Phi^-_{25}
\stackrel{\sigma_1}{\longrightarrow} -\Psi^-_{25}$ and share one certain bit 1 and two random bits 01
privately.Therefore,
 in our proposed protocol, Alice and Bob perform one local operation on their respective particles 1 and 2,
  Charlie  shares 2  certain bits and 2 random bits  with Alice, and 1 certain bit and 2 random bits with Bob secretly.

(B)  Bob can also apply unitary operator $I=|0\rangle\langle 0|+|1\rangle\langle1|$ and ${\rm
i}\sigma_y=|0\rangle\langle1|-|1\rangle\langle0|$,
 and he and Charlie   agree beforehand as the  encoding: $I\rightarrow 0,~~ {\rm i}\sigma_y\rightarrow 1$,
instead of that in the above protocol.

(C) Particles 1 and 2 play symmetric and equal role. That is,
Alice (Bob) can  use one local operation in Eq.(\ref{2operation2})
(Eq.(\ref{2operation1})) and transmit one bit (two bits)
information to Charlie.

(D) There are 4!=24 ( 2!=2 ) kinds of encoding method for one assigning two bits to local operations $I$,
$\sigma_x$, $\texttt{i}\sigma_y$ and $\sigma_z$ ( one bit to $I$ and $\sigma_z$ ). Two communication parties can
choose randomly one kind as their encoding scheme.
\\[0.3cm]
{\noindent{\bf 3. Security  }}\\[0.2cm]
The security of these schemes are limited with the quality of the quantum channel between the parties. We base
our argument of security on perfect quantum channel (that is, the shared GHZ states between the parties are
maximally entangled and free of noise). Since the communication parties  are spatially separated, and one can
not distinguish the noise introduced by the eavesdropper and the noise induced during the preparation and
distribution phases, after generating and distributing such states, the parties may share an ensemble of noisy
GHZ states.  In order to share perfect quantum information channel, they  first purify noisy GHZ states and then
test  the security of quantum channel.  Suppose that the three communication parties share an ensemble of $N'$
identical mixed multi-partite states, they can obtain perfect GHZ states by using efficient multipartite
entanglement distillation protocol---multi-party hashing method \cite{MS} and its improvement \cite{CL}.  After
that, the parties verify if they share perfect maximally entangled GHZ states. They can utilize the similar
method in \cite{Gaozna} to do the tests. In fact, as long as  the states taking as quantum information channel
are the eigenvector of $\sigma_x\otimes \sigma_x\otimes \sigma_x$, $\sigma_z\otimes \sigma_z\otimes I$ and
$\sigma_z\otimes I\otimes \sigma_z$, then the quantum channel is perfect \cite{CL}.

The procedure  obtaining  perfect GHZ states by using efficient multipartite entanglement distillation
protocol---multi-party hashing method \cite{MS} and its improvement \cite{CL} is as follows. Suppose three
parties Alice, Bob, and Charlie share an ensemble of $N'$ identical mixed tripartite states $\rho$ and they
would like to distill out  perfect GHZ states $|P^+\rangle$.  The GHZ state $|P^+\rangle$ is the $+1$ eigenstate
of the following set of commuting observables:
\begin{eqnarray}\label{e:stabilizer}
S_{0}& =&\sigma_x\otimes \sigma_x\otimes \sigma_x, \nonumber \\
S_{1}& =&\sigma_z\otimes \sigma_z\otimes I,  \\
S_{2}& =&\sigma_z\otimes I\otimes \sigma_z. \nonumber
\end{eqnarray}
Denote GHZ states in Eqs.(\ref{2GHZ}) by
\begin{equation}
\left\vert \text{GHZ}_{p,i_{1},i_{2}}\right\rangle _{ABC}=\frac{1}{\sqrt{2}}%
(\left\vert 0\right\rangle \left\vert i_{1}\right\rangle \left\vert
i_{2}\right\rangle +(-1)^{p}\left\vert 1\right\rangle \left\vert \overline{%
i_{1}}\right\rangle \left\vert \overline{i_{2}}\right\rangle) , \label{GHZbasis}
\end{equation}
where $p$ and the $i$'s are zero or one and a bar over a bit value indicates its logical negation. Here, the
three labels $(p,i_{1},i_{2})$ correspond to the eigenvalues of the 3 stabilizer generators $S_{0},S_{1},S_{2}$
by correspondence relation:
\begin{align*}
\text{eigenvalue \ \ }1 & \longrightarrow\text{label }0, \\
\text{eigenvalue}-1 & \longrightarrow\text{label }1.
\end{align*}
According to \cite{dctprl99, dcpra00}, Alice, Bob, and Charlie can depolarize  3-party density matrix $\rho$ by
the following steps. They three perform  the operator $\sigma_x\otimes\sigma_x\otimes\sigma_x$  with a
probability $1/2$, and  then apply $\sigma_z\otimes\sigma_z\otimes I$ with a probability $1/2$. Finally,  they
also apply $\sigma_z\otimes I\otimes\sigma_z$ with a probability $1/2$. The overall operation corresponds to
\begin{equation}
   \begin{array}{lll}
\rho \longrightarrow & \rho_{ABC}= & \frac{1}{8}\big(\rho +(\sigma_x\otimes\sigma_x\otimes\sigma_x)\rho
 (\sigma_x\otimes\sigma_x\otimes\sigma_x)+(\sigma_z\otimes\sigma_z\otimes I)\rho (\sigma_z\otimes\sigma_z\otimes I) \\
&& +(\sigma_y\otimes\sigma_y\otimes\sigma_x)\rho (\sigma_y\otimes\sigma_y\otimes\sigma_x)+(\sigma_z\otimes
I\otimes\sigma_z)\rho (\sigma_z\otimes I\otimes\sigma_z)\\
&& +(\sigma_y\otimes\sigma_x\otimes\sigma_y)\rho (\sigma_y\otimes\sigma_x\otimes\sigma_y)
+(I\otimes\sigma_z\otimes\sigma_z)\rho (I\otimes\sigma_z\otimes\sigma_z)\\
&& +(\sigma_x\otimes\sigma_y\otimes\sigma_y)\rho (\sigma_x\otimes\sigma_y\otimes\sigma_y)\big).
    \end{array}
\end{equation}
The overall operation makes $\rho $ diagonal in the basis Eq.(\ref{GHZbasis})  by the following form:
\begin{eqnarray}\label{e:GHZdiagonal}
\rho_{ABC}=\left(
\begin{array}{cccccccc}
p_{000} & 0 & 0 & 0 & 0 & 0 & 0 & 0 \\
0 & p_{100} & 0 & 0 & 0 & 0 & 0 & 0 \\
0 & 0 & p_{011} & 0 & 0 & 0 & 0 & 0 \\
0 & 0 & 0 & p_{111} & 0 & 0 & 0 & 0 \\
0 & 0 & 0 & 0 & p_{010} & 0 & 0 & 0 \\
0 & 0 & 0 & 0 & 0 & p_{110} & 0 & 0 \\
0 & 0 & 0 & 0 & 0 & 0 & p_{001} & 0 \\
0 & 0 & 0 & 0 & 0 & 0 & 0 & p_{101}%
\end{array}
\right),
\end{eqnarray}
 without changing the diagonal
coefficients.  Thus, three party Alice, Bob, and Charlie share a large ensemble of a density matrix,
$\rho_{ABC}$, that is GHZ-diagonal. They can estimate its matrix elements reliably by using local operations and
classical communications (LOCCs) only.  Measuring along $X, Y, Z$ basis and comparing the results of their local
measurements, they can estimate the diagonal matrix elements in (\ref{e:GHZdiagonal}) by applying classical
random sampling theory. [This is due to the commuting observable argument in \cite{qkd}.]  By definition, any
GHZ-basis vector in Eq.(\ref{GHZbasis})  is a simultaneous eigenvector of the 7 non-trivial stabilizer group
elements $\sigma_x\otimes\sigma_x\otimes\sigma_x$, $\sigma_z\otimes\sigma_z\otimes I$, $\sigma_z\otimes
I\otimes\sigma_z$, $-\sigma_y\otimes\sigma_y\otimes\sigma_x$, $I\otimes\sigma_z\otimes\sigma_z$,
$-\sigma_y\otimes\sigma_x\otimes\sigma_y$, and $-\sigma_x\otimes\sigma_y\otimes\sigma_y$. If the error rates for
all of the 7 non-trivial group elements are denoted $s_1, \ldots, s_7$, then
\begin{align}
p_{000}& =1-\frac{1}{4}(s_{1}+s_{2}+s_{3}+s_{4}+s_{5}+s_{6}+s_{7}), \nonumber \\
p_{100}& =\frac{1}{4}(s_{1}-s_{2}-s_{3}+s_{4}-s_{5}+s_{6}+s_{7}), \nonumber \\
p_{011}& =\frac{1}{4}(-s_{1}+s_{2}+s_{3}+s_{4}-s_{5}+s_{6}-s_{7}), \nonumber \\
p_{111}& =\frac{1}{4}(s_{1}+s_{2}+s_{3}-s_{4}-s_{5}-s_{6}+s_{7}), \nonumber \\
p_{010}& =\frac{1}{4}(-s_{1}+s_{2}-s_{3}+s_{4}+s_{5}-s_{6}+s_{7}), \nonumber \\
p_{110}& =\frac{1}{4}(s_{1}+s_{2}-s_{3}-s_{4}+s_{5}+s_{6}-s_{7}), \nonumber \\
p_{001}& =\frac{1}{4}(-s_{1}-s_{2}+s_{3}-s_{4}+s_{5}+s_{6}+s_{7}), \nonumber \\
p_{101}& =\frac{1}{4}(s_{1}-s_{2}+s_{3}+s_{4}+s_{5}-s_{6}-s_{7}).
\end{align}
Since $s_1$, $\cdots$, $s_7$ can be determined by local operations and classical communications (LOCCs) by
Alice, Bob and Charlie, the above equations relate the diagonal matrix element of the density matrix,
$\rho_{ABC}$, to experimental observables.

 Maneva and Smolin \cite{MS} constructed an efficient multi-partite entanglement distillation protocol---multi-party hashing
 method and showed that its yield (per input
mixed state):
\begin{equation}
D_{h}=1-\max_{j>0}[\{H(b_{j})\}]-H(b_{0}).  \label{smolinhash}
\end{equation}%
  Here $b_{0}$ is formed by concatenating the unknown phase bits of all $\rho_{ABC}$ while $b_{j}$ are
formed by concatenating the $j$th amplitude bits, and
\begin{eqnarray} H(b_{0})
&=&-\sum\limits_{b_{0}=0,1}(\sum%
\limits_{b_{1},b_{2}=0,1}p_{b_{0}b_{1}b_{2}})\log
_{2}(\sum\limits_{b_{1},b_{2}=0,1}p_{b_{0}b_{1}b_{2}}), \nonumber \\
H(b_{1})
&=&-\sum\limits_{b_{1}=0,1}(\sum%
\limits_{b_{0},b_{2}=0,1}p_{b_{0}b_{1}b_{2}})\log
_{2}(\sum\limits_{b_{0},b_{2}=0,1}p_{b_{0}b_{1}b_{2}}),  \\
H(b_{2})
&=&-\sum\limits_{b_{2}=0,1}(\sum%
\limits_{b_{0},b_{1}=0,1}p_{b_{0}b_{1}b_{2}})\log _{2}(\sum\limits_{b_{0},b_{1}=0,1}p_{b_{0}b_{1}b_{2}}).
\nonumber
\end{eqnarray}
Therefore, If $D_h>0$,  using Maneva and Smolin's multi-party hashing
 method, Alice, Bob, and Charlie can distill out $N'D_h$  perfect (generalized)
GHZ states $|P^+\rangle$.  Chen and Lo \cite{CL} presented an improved hashing protocol and proved that its
yield can be increased to:
\begin{equation}
D_{h}^{^{\prime }}=1-\max \{H(b_{1}),H(b_{2}|b_{1})\}-H(b_{0})+I(b_{0};b_{1},b_{2}). \label{newyield}
\end{equation}
With the improved random hashing method of Chen and Lo, Alice, Bob, and Charlie  can distill out
$N'D_{h}^{^{\prime }}$ perfect (generalized) GHZ states $|P^+\rangle$ if $D_{h}^{^{\prime }}>0$.

The only place an eavesdropping can affect the system is the distribution phase of the GHZ states between the
communication parties. If  the eavesdropper couples her ancilla states during preparing or distribution GHZ
state, the communication parties can find out her by the method of \cite{Gaozna}, and remove the entanglement
between eavesdropper's particles and GHZ tripartite by multi-party hashing
 method \cite{MS}.  That is, by testing the security of quantum channel, the eavesdropper can be detected,
 and as long as $D_{h}>0$ ($D_{h}^{^{\prime }}>0$), the three communication parties can get perfect GHZ states. However,
by testing the security of quantum channel, if $D_{h}=0$ ($D_{h}^{^{\prime }}=0$), Alice, Bob and Charlie
discard the quantum channel and construct it again.
 In one word, under any case, as long as an eavesdropper exists, we can find her and ensure the safety of
quantum channel.   Once the security of the quantum channel is assured, which means that  Alice, Bob and Charlie
share pure GHZ triplets (perfect quantum channel), then no information is leaked to Eve. Hence
 our proposed protocol is secure, even if the shared quantum channels are
public.\\[0.2cm]
{\noindent{\bf 4. Summary }}\\[0.2cm]
We present a new deterministic secure method for direct communication  by GHZ states and swapping quantum
entanglement, where the three  spatially separated parties  faithfully transmit secret messages  and detect
eavesdroppers by the correlations of entanglement swapping results. In our scheme the secret messages can be
encoded directly and are faithfully transmitted from two senders  Alice and Bob  to a remote receiver Charlie at
the same time via initially shared GHZ states without revealing any information to a potential eavesdropper. The
distributed entangled particles shared by Alice, Bob,   and Charlie function as a quantum information channel
for faithful transmission. Using $2N$  GHZ states, Alice can send $2N$ bits secret messages to Charlie,
meanwhile, Bob can also transmit $N$ bits information to Charlie. Since there is not a transmission of the qubit
carrying the secret message between Alice and Bob, and Charlie in the public channel, it is completely secure
for direct secret communication if perfect quantum channel is used.  That is, simultaneous many mutual QSDC of
 the central party and other two parties can be realized.
\\[0.2cm]
{\bf Acknowledgments}\\[0.2cm]  This work was supported by Hebei Natural Science Foundation  under Grant
 No. A2004000141 and No. A2005000140, and Natural Science
Foundation of Hebei Normal University of China.

\end{document}